\documentclass[letterpaper, 10 pt, conference]{ieeeconf}  

\IEEEoverridecommandlockouts                              
\usepackage{anysize}
\usepackage{algorithmicx,algorithm}
\usepackage{amsfonts}
\usepackage{amssymb,amsmath}
\usepackage{booktabs} 
\usepackage{color}
\usepackage{epsf}
\usepackage{epsfig}
\usepackage{float}
\usepackage{graphicx}
\usepackage{hyperref}
\usepackage{mathrsfs}
\usepackage{microtype}
\usepackage{subfigure}
\usepackage{siunitx}
\usepackage[usenames,dvipsnames]{xcolor}
\usepackage{algpseudocode}

\usepackage{amsmath,amsfonts,amssymb}
\usepackage{mathtools}
\usepackage{bm}
\usepackage{braket}
\usepackage{mathrsfs}
\usepackage{theorem}
\usepackage{multirow}

\newtheorem{definition}{Definition}
\newtheorem{proposition}{Proposition}

\newtheorem{theorem}{Theorem}

\newcommand{\dd}{{\rm d}}

\title{\LARGE \bf
On the Role of Controllability in Pulse-based Quantum Machine Learning Models
}

\author{Han-Xiao Tao and Re-Bing Wu
\thanks{This work was supported by NSFC (Grant No. 62173201).}
\thanks{HXT and RBW are with the Center for Intelligent and Networked Systems, Department of Automation, Tsinghua University,
        Beijing, 100084, China.
        {\tt\small rbwu@tsinghua.edu.cn}}%
}

\begin{document}

\maketitle
\thispagestyle{empty}
\pagestyle{empty}

\begin{abstract}

Pulse-based quantum machine learning (QML) models possess full expressivity when they are ensemble controllable. However, it has also been shown that barren plateaus emerge in such models, rendering training intractable for systems with large dimension. In this paper, we show that the trade-off is closely related to the controllability of the underlying pulse-based models. We first apply the Fliess-series expansion to pulse-based QML models to investigate the effect of control system structure on model expressivity, which leads to a universal criterion for assessing the expressivity of generic QML models. Guided by this criterion, we then demonstrate how designing pulse-based models on low-dimensional manifolds can balance expressivity and trainability. Finally, numerical experiments are carried out to verify the proposed criterion and our analysis, which futher demonstrate that increasing dimensionality enhances expressivity but avoids barren plateaus if the model is designed with limited controllability on a submanifold. Our approach provides a promising path for designing pulse-based QML models that are both highly expressive and trainable.

\end{abstract}

\section{INTRODUCTION}

Machine learning powered by fault-tolerant quantum computers can substantially improve the training or inference performance~\cite{arute2019quantum,zhong2020quantum,zhu2022quantum,huang2022quantum}. However, in the so-called noisy intermediate-scale quantum (NISQ) era, quantum machine learning (QML) has to rely on variational models due to the unavailability of error-correctible qubits~\cite{endo2021hybrid,callison2022hybrid}. Such variational QML models predominantly employ parametrized quantum circuits, also referred to as quantum neural networks (QNNs)~\cite{farhi2018classification,mcclean2018barren}, with gate parameters optimized through external classical algorithms. While state-of-the-art NISQ devices can implement QML models intractable for classical simulation, hardware restrictions on coherent evolution time limit realizable circuit depth. Consequently, model expressivity is impeded owing to the linear scaling between trainable parameters and circuit depth.

From a hardware perspective, QML models may employ continuous control pulses rather than discrete gate parameters for parameterization. In contrast to gate-based models, pulse-based models do not require pre-selection of a circuit ansatz at the abstraction level. Owing to the proximity to the underlying physics, pulse-based models are friendly to hardware engineers as they permit simpler design and implementation~\cite{wu2020end,pan2023experimental,liang2022variational,melo2023pulse,ibrahim2022pulse}. Furthermore, given fixed coherence time, pulse-based QML models can unleash more expressive power compared to the gate-based counterparts through realization of effectively "infinitely" deep QNNs~\cite{tao2024unleashing}. 

Numerous investigations have been casted to the expressivity of QML models, but predominately from the perspective of gate-based architectures, e.g., the ability to uniformly traverse the Hilbert space~\cite{sim2019expressibility,du2020expressive,morales2020universality,biamonte2021universal,chen2021universal}, the pseudo-dimension~\cite{caro2020pseudo}, the covering number~\cite{du2022efficient}, the Rademacher complexity~\cite{bu2022statistical}, and the memory capacity~\cite{wright2020capacity}. The expressivity can also be understood from the ability of approximating nonlinear functions, including the kernel-based perspective~\cite{havlivcek2019supervised,schuld2019quantum,schuld2021supervised}, the data-replication and data-reuploading~\cite{goto2021universal,wu2021expressivity,PerezSalinas2020datareuploading,PhysRevA.104.012405,schuld2021effect,gan2022fock}. 

Regarding the expressivity of pulse-based QML models, we showed in our previous work that ensemble controllable models have the capability of approximating arbitrary functions~\cite{tao2024unleashing}. However, the full controllability inevitably induce barren plateaus in the loss landscape given sufficient evolution time~\cite{larocca2022diagnosing,banchi2017driven}. This phenomenon causes gradients to vanish exponentially with qubit number, rendering models intractable to optimize for large systems~\cite{mcclean2018barren,cerezo2021cost,wang2021noise,arrasmith2021effect,cerezo2021higher,holmes2021barren,sharma2022trainability}. Research on gate-based QML has also demonstrated that stronger model expressivity leads to smaller gradients and poor trainability~\cite{holmes2022connecting}. These findings reveal an inherent trade-off between model expressivity and trainability that must be balanced. 

In this paper, we will analysis the expressivity of pulse-based QML models through Fliess series expansions and formulate a universal theorem for evaluating the expressivity of pulse-based models. Leveraging the proposed theorem, we construct models on a low-dimensional submanifold of the system to simultaneously achieve the expressive power for arbitrary functions along with excellent optimization properties that can avoid barren plateaus. 

The remainder of this paper is arranged as follows. Section~\ref{PULSE-BASED QML MODELS} gives necessary background about pulse-based QML models. Section~\ref{EXPRESSIVITY OF QML MODELS} begins by presenting our previous universal expressivity criterion. Subsequently, it explains the expressivity of pulse-based models from the perspective of Fliess series expansion, based on which a novel theorem is introduced to quantify the expressive power of the models. In Section~\ref{TRAINABILITY OF QML MODELS}, we discuss the trainability of QML models and introduce the proposition that the full controllability of the pulse-based model leads to the emergence of barren plateaus. In Section~\ref{NUMERICAL EXPERIMENTS}, numerical experiments on the expressivity of pulse-based models and the method for balancing model expressivity and trainability are demonstrated. Finally, conclusion and outlooks are made in Section~\ref{CONCLUSION AND OUTLOOK}.

\section{PULSE-BASED QML MODELS}
\label{PULSE-BASED QML MODELS}

This section provides an introduction to the pulse-based model and its primary universal approximation theorem, with further details available in our previous work~\cite{tao2024unleashing}.

Different from gate-based models, pulse-based models is parameterized by continuous-time control pulses instead of discrete-time gate parameters. The nonlinearity of pulse-based models arises from the encoding process, which can be regarded as the continuous limit of data reuploading in gate-based models. 

In general, pulse-based models can be described as the following type of controlled quantum systems:
\begin{equation}
\label{quantum system}
\ket{\dot\psi(t;\textbf{x})}=-i\left[H_0(\textbf{x})+\sum_{k=1}^p\theta_k(t)H_{k}\right]\ket{\psi(t;\textbf{x})},
\end{equation}
that is parameterized by continuous-time control
functions $\theta_k(t)(k=1,\cdots,p)$, where $\ket{\psi(0)}=\ket{\psi_0}$, and the input variable $\textbf{x}=(x_1,\cdots,x_m)$, belonging to some domain $\mathcal{X}\subseteq \mathbb{R}^m$, is uploaded via the Hamiltonian 
\[
H_0(\textbf{x})=x_1D_1+\cdots+x_mD_m
\] with $D_1,\cdots,D_m$ being the corresponding data-encoding Hamiltonians. Similar to gate-based models, the output is inferred via the expectation-value measurement
\begin{equation}\label{eq:measurement}
  f_\Theta(\textbf{x})=\langle \psi(T;\textbf{x})|M|\psi(T;\textbf{x})\rangle,
\end{equation}
of some observable $M$ to approximate multi-variable functions through optimization of the control function $\Theta(t)=(\theta_1(t),\cdots,\theta_p(t))$. This model can be physically realized by a superconducting quantum system, where the microwave Rabi control pulse $\theta_k(t)(k=1,\cdots,p)$ induces coherent transition between the qubit's two logical states. Since the microwave control pulse for superconducting qubit control is generated by an Arbitrary Waveform Generator in the form of piecewise-constant pulses, the untiary transformations generated the sub-pulses can be regarded as the QNN layers.

Compared to gate-based models, pulse-based models are more hardware efficient because the hardware qubit-connectivity topology can be directly seen from the Hamiltonian structure. Besides, pulse-based models can be designed deeper within the same coherence time. Therefore, pulse-based models exhibit greater expressivity relative to gate-based models implemented on equivalent hardware architectures.

\section{EXPRESSIVITY OF QML MODELS}
\label{EXPRESSIVITY OF QML MODELS}
It was found that pulse-based models can approximate arbitrary nonlinear functions, i.e. have full expressivity, when the underlying quantum control system is ensemble controllable. This universal expressivity criterion has been summarized in the following theorem~\cite{tao2024unleashing}.
\begin{theorem}\label{thm:universal approximation properties_0}
The pulse-based model~\eqref{quantum system} can approximate any function $f: \mathcal{X}\rightarrow[\lambda_{\min},\lambda_{\max}]$ if the Lie algebra $\{iH_{1},\cdots,iH_{p}\}_{LA}=su(d)$.
\end{theorem}

However, the above criteria established for expressivity is a rather strong condition. Moreover, as will be discussed in Section~\ref{TRAINABILITY OF QML MODELS}, full controllability lead to barren plateaus, rendering training intractable for large-scale models. Consequently, there is a necessary to investigate a weaker condition for verified model expressivity and balance expressivity versus trainability.

In this section, we will show that the expressivity of the pulse-based QML model can also be analyzed based on Fliess series expansion. We refer the reader to~\cite{lamnabhi2009volterra,isidori1995nonlinear} for a detailed introduction to Fliess series,

For simplicity, we consider the univariate case with the following Hamiltonian:
\begin{equation}
\label{quantum system2}
H\left[x;\Theta(t)\right]=xH_0+\sum_{j=1}^p\theta_j(t)H_{j}.
\end{equation}
Let $\mathcal{L}_j$ be the Liouvillian such that $\mathcal{L}_j M\triangleq [-iH_j,M]$ and $\mathbb{I}_k$ be the collection of tuples $(j_1,\cdots,j_n)$ ($n\geq k$) of which $k$ indices are 0. Denote $\theta_0(t)\equiv 1$, then we can express the function \eqref{eq:measurement} as a power series using the Fliess expansion: 
\begin{equation}
\label{power series}
f_\Theta(x)= \sum_{k=0}^\infty C_kx^k,    
\end{equation}
 where
\begin{equation}\label{eq:coefficient}
C_k=\sum_{(i_1,\cdots,i_n)\in \mathbb{I}_k}c_{j_1\cdots j_n}(T)\langle\psi_0|\mathcal{L}_{j_1}\cdots\mathcal{L}_{j_n}M|\psi_0\rangle, 
\end{equation}
where
\begin{equation}
\begin{split}
 c_{j_1\cdots j_n}(T)= &\int_0^T\theta_{j_1}(t_1)\dd t_1\int_0^{t_1}\theta_{j_2}(t_2)\dd t_2 \\
  &\cdots\int_0^{t_{n-1}}\theta_{j_n}(t_n)\dd t_n.
\end{split}
\end{equation}
Given that for any smooth function $f(x)$, the Taylor series expansion around the point $x=0$ produces a polynomial in $x$ that can approximate the original function $f(x)$. Therefore, if arbitrary set of coefficients $\{c_k\}$ can be generated through some $\Theta(t)$, then the control system has the ability of approximating arbitrary analytic function as long as $\lambda_{\min} \leq f_\Theta(x) \leq \lambda_{\max}$, where $\lambda_{\min}$ and $\lambda_{\max}$ are the smallest and largest eigenvalues of $M$. 

To facilitate the analysis, we define the following set
\[
\mathscr{O}(X)={\rm span}\{\mathcal{L}_{j_1}\cdots\mathcal{L}_{j_n}X,~~1\leq j_k\leq p,~n\in\mathbb{N}.\}.
\]
Let $\mathcal{S}_0=\mathscr{O}(M)$ and 
\[\mathcal{S}_k=\mathscr{O}\left(\mathcal{L}_0 \mathcal{S}_{k-1}\right).\]

Let $\mathcal{S}$ be the set of all operators appeared in \eqref{power series}. It is easy to prove that 
\[\mathcal{S}=\bigcup_{k=0}^\infty \mathcal{S}_k.\]
Consider the class of functions that can be Taylor expanded in a neighborhood of $x=0$. Then we can form the following theorem.

\begin{theorem}
\label{thm_1}
The pulse-based QML model~\eqref{quantum system} cannot approximate any function $f: \mathcal{X}\rightarrow[\lambda_{\min},\lambda_{\max}]$ if there exists some $k$ such that $\bra{\psi_0}\mathcal{S}_k \ket{\psi_0}=\{0\}$.
\end{theorem}

In other words, the condition $\bra{\psi_0}\mathcal{S}_k \ket{\psi_0}\neq\{0\}$ for all $k\geq 0$ is a necessary condition for the expressivity of the QML model, but whether it is a sufficient condition needs further investigation.

For example, consider the QML model with the following Hamiltonian
\begin{equation}
\label{eq:example_2}
H[x;\theta(t)]=x\sigma_{\rm z}^{(1)}\sigma_{\rm z}^{(2)}+\theta_1(t)\sigma_{\rm x}^{(1)}+\theta_2(t)\sigma_{\rm y}^{(1)}
\end{equation}
where $\sigma_\alpha^{(k)}=\mathbb{I}_2\otimes\cdots\otimes\sigma_\alpha\otimes\cdots\otimes \mathbb{I}_2$ with $\sigma_\alpha$ on the $k$th site, and the observable is chosen as $M=\sigma_{\rm z}^{(1)}\sigma_{\rm z}^{(2)}$. Using the above theorem, we can obtain
\begin{eqnarray*}
\mathcal{S}_{2k+1}&=&\{\sigma_{\rm x}^{(1)},\sigma_{\rm y}^{(1)},\sigma_{\rm z}^{(1)}\},\\
\mathcal{S}_{2k}&=&\{\sigma_{\rm x}^{(1)}\sigma_{\rm z}^{(2)},\sigma_{\rm y}^{(1)}\sigma_{\rm z}^{(2)},\sigma_{\rm z}^{(1)}\sigma_{\rm z}^{(2)}\},
\end{eqnarray*}
for $k=0,1,2,\cdots$. When the initial state $|\psi_0\rangle=|0\rangle \otimes |0\rangle$, we can examine that $\bra{\psi_0}\mathcal{S}_k \ket{\psi_0}\neq \{0\}$ for $\forall k$, so the model satisfies the necessary condition in Theorem~\ref{thm_1}, and we can numerically verify it has the ability to express any function in Section~\ref{NUMERICAL EXPERIMENT_1}. However, if the initial state $|\psi_0\rangle=|0\rangle \otimes (\frac{1}{\sqrt{2}}|0\rangle+\frac{1}{\sqrt{2}}|1\rangle)$, then $\bra{\psi_0}\mathcal{S}_{2k} \ket{\psi_0}= \{0\}, \bra{\psi_0}\mathcal{S}_{2k+1} \ket{\psi_0}\neq \{0\}$ for $\forall k$, which indicates that all even power terms of $x$ vanishes, and we will demonstrate in Section~\ref{NUMERICAL EXPERIMENT_1} that only odd functions can be approximated under this circumstance.

\section{TRAINABILITY OF QML MODELS}
\label{TRAINABILITY OF QML MODELS}
Trainability is another inherent capability of the learning model, which denotes its capacity to efficiently undergo iterative optimization procedures, adapting its parameters in response to training data to converge towards optimal performance. In the field of QML, a notorious phenomenon known as the barren plateau can lead to significantly poor trainability of models. 

In brief, when a cost function exhibits a barren plateau, the gradients exponentially vanish throughout the optimization landscape with the increasing number of qubits. This suggests that achieving effective training for models with a large number of qubits is nearly impractical due to the requirement for an exponentially high precision to navigate the flat landscape and ascertain a direction that minimizes cost~\cite{mcclean2018barren,cerezo2021cost}. The barren plateau phenomenon is defined as follows~\cite{larocca2022diagnosing}:

\begin{definition}\label{definition:BP} 
The cost function $L[\Theta(t)]$ for the model~\eqref{quantum system} with $n$ qubits is said to have a barren plateau if the variation
of partial derivative satisfys
\[
{\rm Var}\left[\frac{\partial L}{\partial\Theta(t)}\right]\leq F(n), 
\]
for all $t>0$, where $F(n) \in \mathcal{O}(b^{-n})$ for some $b>1$. 
\end{definition}

It has been proven that fully controllable pulse-based models always converge to unitary $t$-designs in the long-time limit~\cite{banchi2017driven}, which are known to have barren plateaus~\cite{mcclean2018barren}. The following proposition articulates the conditions for the exhibition of barren plateaus in terms of controllability~\cite{larocca2022diagnosing}.
\begin{proposition}\label{proposition:BP} 
There exists a scaling of the depth for which controllable systems form $\epsilon$-approximate $2$-designs with $\epsilon \in \mathcal{O}\left(1/2^n\right)$, and hence
the system exhibits a barren plateau.
\end{proposition}

This proposition indicates that barren plateaus will arise once the QNN is sufficiently deep so that the underlying system is controllable. Therefore, pulse-based models that are expressive but with limited controllability may potentially avoid the occurrence of barren plateaus, thereby enhancing the trainability of the model.

\section{NUMERICAL EXPERIMENTS}
\label{NUMERICAL EXPERIMENTS}
This section will demonstrate the expressive power of pulse-based QML models and the balance between expressivity and trainability via numerical experiments. The observable $M$ is always chosen with $\lambda_{\min}=-1$ and $\lambda_{\max}=1$. For the sake of uniformity, we may set the domain of $\textbf{x}$ as $\mathcal{X}=[-1,1]^m$ without loss of generality~\cite{tao2024unleashing}.
 
In the training of the pulse-based QML model, the duration $T$ and the number $K$ of discrete subpulses need to be preselected as hyperparameters. The parameter $K$ represents the number of layers in QNN models, while the length of sampling period $\Delta t=T/K$ determines to what extent the quantum state can be transformed from one layer to the next layer. The QML model is trained by minimizing the mean squared error (MSE) 
\begin{equation}
\label{cost_func}
L[\Theta(t)]=N^{-1}\sum_{k=1}^N\left\|f_{\Theta}(\textbf{x}^{(k)})-y^{(k)}\right\|^2
\end{equation}
over the training dataset $\{(\textbf{x}^{(k)},y^{(k)})\}$ sampled from the target function to be fitted. Here the Adam optimizer is used with a learning rate of 0.1. All simulations are carried out with Matlab, using a server with two 10-core Xeon CPUs and 208GB RAM.

\subsection{Approximation of Nonlinear Functions via Pulse-based QML Models}
\label{NUMERICAL EXPERIMENT_1}

\begin{figure}[thpb]
\centering
\includegraphics[width=1\columnwidth]{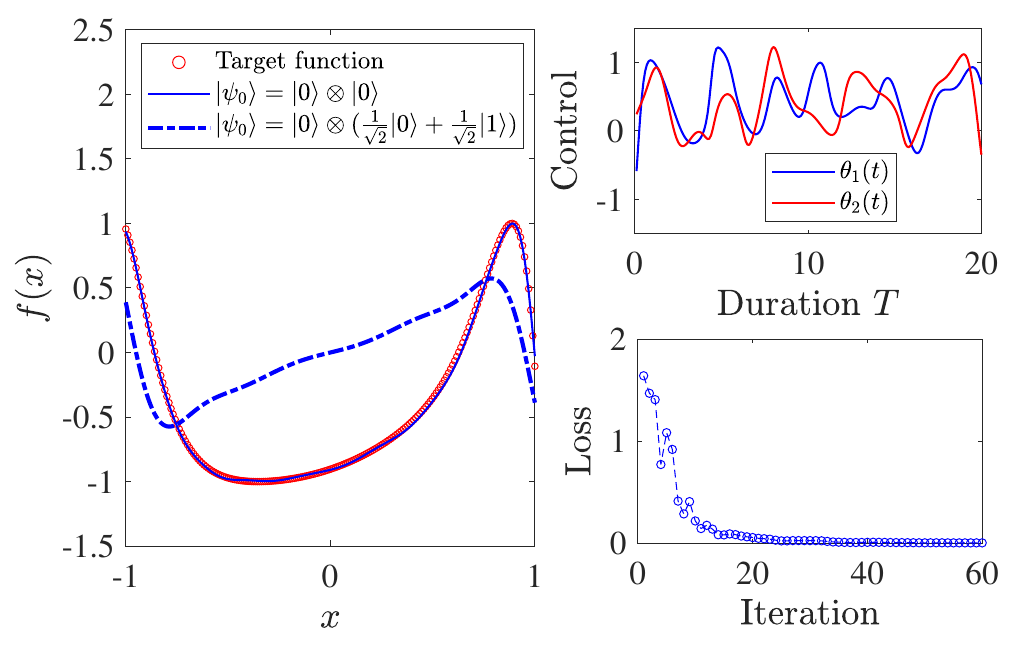}\hfill
\caption{The approximation results of the polynomial function by the two-qubit models. On the left are the fitted curve, and on the right are the resulting control fields and the training curves.}
\label{Fig:fitting}
\end{figure}

We first apply the model~\eqref{eq:example_2} to approximate the following target function:
\begin{small}
\begin{equation}
\label{poly}
  f_{1}(x)=2x+3x^2+x^3+10x^6+8x^7-3x^9+5x^{10}-13x^{12},
\end{equation}
\end{small}
from which $200$ points are evenly sampled to train the learning model. We set $\Delta t=0.1$ and $K=200$. The simulation result shown in Fig.~\ref{Fig:fitting}(b) indicates that the trained QML model can fit the target function pretty well after $60$ rounds of iteration when $|\psi_0\rangle=|0\rangle \otimes |0\rangle$. However, the model exhibits limited expressive power when the initial state is selected as $|\psi_0\rangle=|0\rangle \otimes (\frac{1}{\sqrt{2}}|0\rangle+\frac{1}{\sqrt{2}}|1\rangle)$, because in this configuration the model can only approximate odd functions, which validates the analysis in Section~\ref{EXPRESSIVITY OF QML MODELS}.

\subsection{Balance Between Expressivity and Trainability}
\label{NUMERICAL EXPERIMENT_2}
As analyzed in Sections \ref{EXPRESSIVITY OF QML MODELS} and \ref{TRAINABILITY OF QML MODELS}, pulse-based QML models can achieve full expressivity when they are ensemble controllable. However, in such cases, it can lead to the emergence of barren plateaus, rendering the models nearly untrainable for large systems. Thus, achieving a balance between expressivity and trainability represents a critical challenge in the field of QML. In the following, we demonstrate constructing models capable of resolving this challenge by leveraging Theorem~\ref{thm_1}.

Notably, Theorem~\ref{thm_1} indicates that pulse-based models satisfying the necessary condition for expressivity may achieve full expressivity without ensemble controllability. Therefore, the conflict between expressivity and trainability in QML models can be reconciled by designing pulse-based models within low-dimensional subspaces that meet the requirement for expressivity in Theorem~\ref{thm_1}. For instance, we can design the model Hamiltonians within $SU(2)$ subspace of a high-dimensional system. For a $d$-dimensional system, we construct the following model Hamiltonian:
\begin{equation}
\label{eq:example_su2}
  H(t)= xJ_{\rm z}+\theta_1(t)J_{\rm x}+\theta_2(t)J_{\rm y},
\end{equation}
where $\{J_{\rm x},J_{\rm y},J_{\rm z}\}_{LA}=su(2)$ is the $d$-dimensional unitary irreducible representation of $SU(2)$, and the observable is chosen as $\sigma_{\rm z}^{(1)}$. If the initial state $|\psi_0\rangle=|0\rangle \otimes |0\rangle$, we can verify that the model fulfills the necessary condition for expressivity and validate its expressivity numerically. On the other hand, the model is uncontrollable over the entire $d$-dimensional space in this setting. Thus, this pulse-based model possesses both the capability to express arbitrary functions as well as avoidance of barren plateaus. We will further illustrate this through simulation experiments.

\begin{figure}[thpb]
\centering
\includegraphics[width=1\columnwidth]{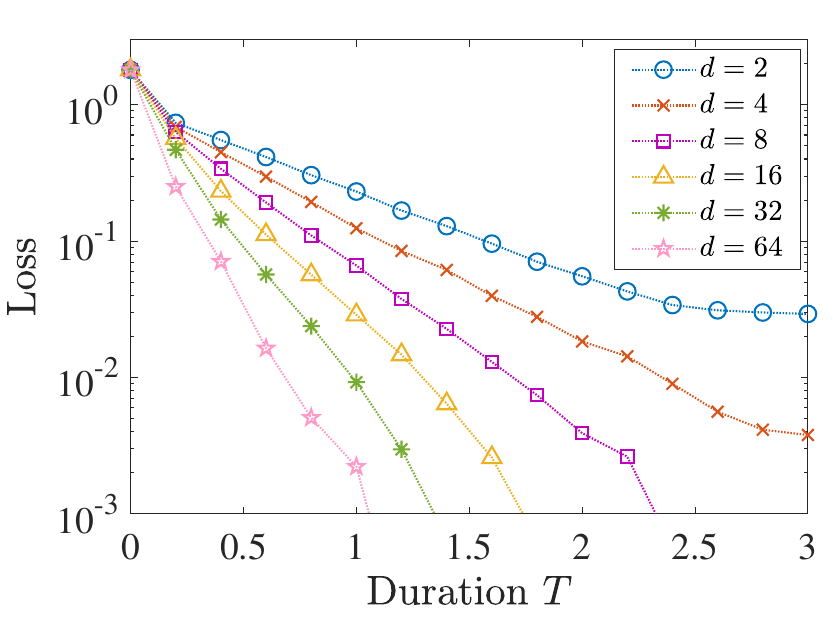}\hfill
\caption{The dependence of training loss on the dimension of control system with the model~\eqref{eq:example_su2}.}
\label{Fig:T-Loss-su2}
\end{figure}

We fit the sigmoid-like function
\begin{eqnarray}
\label{sigmoid}
 f_{2}(x) &=& \frac{1-e^{-10x}}{1+e^{-10x}}
\end{eqnarray}
using the model~\eqref{eq:example_su2} with up to six qubits, where the operators $J_x,J_y,J_z$ are endowed with the corresponding unitary irreducible representation of $su(2)$. Their training-loss curves are plotted in Fig.~\ref{Fig:T-Loss-su2} versus the pulse duration $T$, where the sampling period is fixed as $\Delta t=0.1$.

The results presented in Fig.~\ref{Fig:T-Loss-su2} clearly demonstrate that all models are capable of accurately approximating the sigmoid function, consistent with the above theoretical analysis. Furthermore, the utilization of higher dimensional Hilbert spaces through increased qubit number is shown to shorten the required pulse duration with the same approximation error. This indicates that the expressivity of the QML models strengthens with greater dimensionality of the quantum system.

\begin{figure}[thpb]
\centering
\includegraphics[width=1\columnwidth]{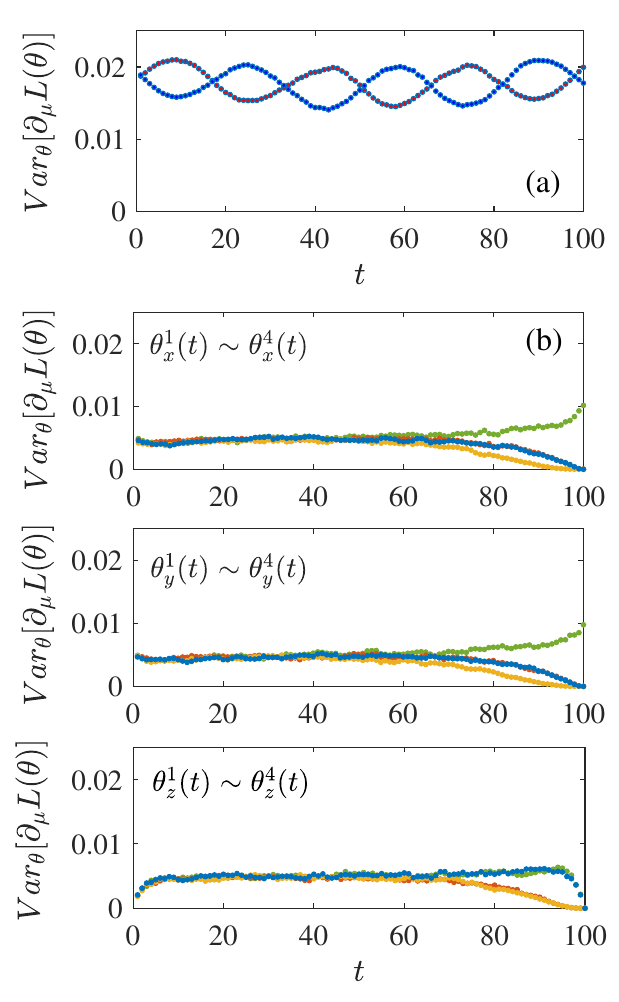}\hfill
\caption{The sample variance of cost function partial derivatives for all parameters of (a) uncontrollable model and (b) controllable model.}
\label{Fig:var-para}
\end{figure}

To compare the trainability of the uncontrollable model~\eqref{eq:example_su2} and the ensemble controllable model, we investigate the sampling variance of the gradients of the cost function~\eqref{cost_func}. Here, the Hamiltonian of the ensemble controllable model we choose is the following circularly coupled $n$-qubit system in~\cite{tao2024unleashing}:
\begin{small}
\begin{equation}\label{eq:simulation}
\begin{split}
&H[x;\Theta(t)]=x\sum_{k=1}^n\sigma_{\rm z}^{(k)}+\sum_{k=1}^n \left[\theta_x^{(k)}(t)\sigma_{\rm x}^{(k)}
+\theta_y^{(k)}(t)\sigma_{\rm y}^{(k)}\right] \\
&+\theta_z^{(1)}(t)\sigma_{\rm z}^{(1)}\sigma_{\rm z}^{(2)}+\cdots+\theta_z^{(n)}(t)\sigma_{\rm z}^{(n)}\sigma_{\rm z}^{(1)}.
\end{split}
\end{equation}
\end{small}
The sample variance of cost function partial derivatives for all parameters of these two models is shown in Fig.~\ref{Fig:var-para}, where we set the qubit number $n=4$, the sampling period $\Delta t=0.1$, and the duration $T=10$. It is evident that the controllable model demonstrates overall lower variance of the parameter gradients, with some approaching 0. Since the variances of the gradients for all parameters remain in close proximity, we subsequently focus our analysis on the gradient of the cost function with respect to the first parameter of the model.

\begin{figure}[thpb]
\centering
\includegraphics[width=1\columnwidth]{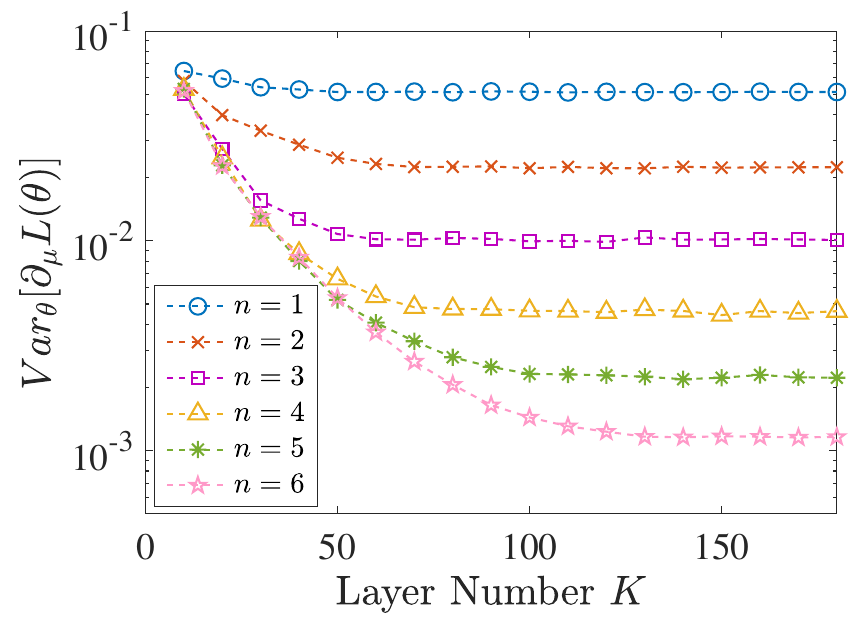}\hfill
\caption{The dependence of the sample variance of the gradient on the layer number of the controllable model.}
\label{Fig:Var-K}
\end{figure}

As illustrated in Fig.~\ref{Fig:Var-K}, the variance of the gradient for the controllable model decreases with the increase of the number $K$ of discrete subpulses. When $K$ reaches a certain threshold, the variance tends to be stable. This threshold increases as the number of qubits grows. The convergence of variance indicates that QNN has converged to $2$-design, where the model starts to exhibit barren plateaus, consistent with the results shown in~\cite{mcclean2018barren}.

\begin{figure}[thpb]
\centering
\includegraphics[width=1\columnwidth]{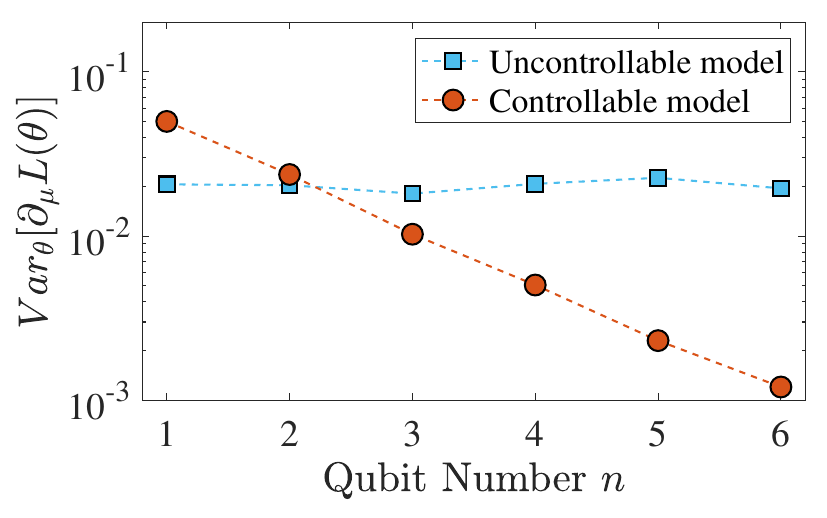}\hfill
\caption{The sample variance of cost function partial derivatives versus qubit number for the controllable and uncontrollable systems.}
\label{Fig:Var}
\end{figure}

In order to demonstrate the trainability advantages of the model~\eqref{eq:example_su2} compared to ensemble controllable models, we present the sample variance of cost function partial derivatives as a function of qubit number $n$ for the cost function~\eqref{cost_func} in a log-linear scale in Fig.~\ref{Fig:Var}, where we set the sampling period $\Delta t=0.1$, and the duration $T=200$ which is sufficient long according to the result in Fig.~\ref{Fig:Var-K}. Here, the variance is computed by randomly initializing $1000$ sets of parameters.

The QML model with controllable system will exhibit barren plateaus, which means the variance of cost function partial derivatives vanishes exponentially with the system size in the long-time~\cite{larocca2022diagnosing}. In Fig.~\ref{Fig:Var} we see that, as expected, the variance for the controllable case is polynomially vanishing with the qubit number $n$. In contrast, the variance remains relatively constant for the uncontrollable case, which implies that the model~\eqref{eq:example_su2} can avoid the emergence of barren plateaus. The above results validate that constructing pulse-based models within a low-dimensional submanifold of high-dimensional quantum systems enables simultaneously achieving strong expressive power while mitigating barren plateaus. 

\section{CONCLUSION AND OUTLOOK}
\label{CONCLUSION AND OUTLOOK}
In summary, we have explored the expressive power of pulse-based QML models from the perspective of Fliess series expansion, and investigated the trade-off between model expressivity and trainability. We propose a criterion by which the expressivity of pulse-based QML models can be assessed. Based on this theorem, we construct pulse-based models on low-dimensional submanifolds that simultaneously exhibit full expressivity while avoiding barren plateaus, achieving the balance between expressivity and trainability. Numerical simulations further demonstrate enhanced expressivity with increasing system dimension for such uncontrollable models designed on submanifolds, while maintain good trainability.

It should be noted that the verification process of Theorem \ref{thm_1} becomes challenging when the model is relatively complex, which requires intricate Lie algebra calculations. Therefore, in subsequent work, we intend to develop a theorem, based on Theorem \ref{thm_1}, that is easy to verify, even for individuals who are unfamiliar with Lie algebra tools. Moreover, the sufficiency of the condition for the expressivity of the pulse-based model in Theorem \ref{thm_1} also warrants further investigation.

This work provides a novel approach for designing pulse-based QML models with balanced expressivity and trainability, alleviating the dilemma of simultaneously achieving both. However, possessing strong expressive power does not necessarily imply good performance on previously unseen data, which requires considering another key property of learning models - generalization. In the QML field, balancing expressive power and generalization premains an intricate issue, as enhanced model expressivity often leads to poorer generalization. In future work, leveraging the proposed pulse-based framework and tools from Lie algebra, we expect to investigate the generalization of QML models and search for architectures with a felicitous combination of expressivity, trainability, and generalization.

\addtolength{\textheight}{-3cm}   
                                  
\section{ACKNOWLEDGMENTS}

The authors acknowledge the support of National Science Foundation of China under grant 62173201.



\end{document}